\documentstyle[12pt]{article}
\newcommand{\extraspace}{\addtolength{\abovedisplayskip}{2mm}
                        \addtolength{\belowdisplayskip}{2mm}
                        \addtolength{\abovedisplayshortskip}{2mm}
                        \addtolength{\belowdisplayshortskip}{2mm}}
\newcommand{\be}{\begin{equation}\extraspace}

\newcommand{\ee}{\end{equation}}

\newcommand{\bea}{\begin{eqnarray}\extraspace}
\newcommand{\beastar}{\begin{eqnarray*}\extraspace}
\newcommand{\eea}{\end{eqnarray}}
\newcommand{\eeastar}{\end{eqnarray*}}

\newcommand{\nonu}{\nonumber \\[2mm]}

\newcommand{\str}{\rule[-2.5mm]{0mm}{7mm}}
\newcommand{\Str}{\rule[-3.5mm]{0mm}{8mm}}
\newcommand{\STr}{\rule[-4.5mm]{0mm}{11mm}}
\newcommand{\STR}{\rule[-5.5mm]{0mm}{13mm}}

\def\xvec{(x_{_1},\ldots,x_{_{N_e}})}

\def\yvec{y_{_1},\ldots,y_{_M}}

\newcommand{\half}{\frac{1}{2}}

\newcommand{\si}{\sigma}
\newcommand{\La}{\Lambda}
\newcommand{\de}{\delta}

\newcommand{\np}{Nucl.Phys.\ }

\begin{document}

%

\baselineskip=12pt

\hfill {ITP-SB-91-51}
\vskip .1cm
\hfill { Nov.  1991}
\vskip 2.0cm
\begin{center}

{\LARGE Completeness of the SO(4) extended }\\
\vspace{5mm}
{\LARGE Bethe Ansatz}\\
\vspace{5mm}
{\LARGE for the One-dimensional Hubbard Model}

\vskip 2.0cm

{\large Fabian H.L. E\char'31 ler}\\
\vspace{3mm}
{\large Vladimir E. Korepin}\\
\vspace{3mm}
{\large  and}\\
\vspace{3mm}
{\large Kareljan Schoutens}

\vskip .5cm

\baselineskip=15pt

{\sl Institute for Theoretical Physics\\
     State University of New York at Stony Brook\\
     Stony Brook, NY 11794-3840, U.S.A.}

\vskip 2.0cm

{\bf Abstract}

\end{center}

\vspace{.5cm}

\noindent
\baselineskip=17pt
We show how to construct a complete set of eigenstates of
the hamiltonian of the one-dimensional Hubbard model on a lattice of
even length $L$.  This is done by using the nested Bethe Ansatz {\it and}
the $SO(4)$ symmetry of the model. We discuss in detail how the counting of
independent eigenstates is carried out.

\vfill


\newpage

\baselineskip=17pt

\section{Introduction}
\setcounter{section}{1}

An important ingredient in the search for a theory of high-$T_c$
superconductors is the analysis of strongly correlated electron
systems $^{\hbox{\scriptsize\cite{and}}}$.
A prototype model for these is the Hubbard model.
Especially interesting is the $\eta$-pairing mechanism proposed by
C.N.Yang in {\cite{ya,yazh,yanew}}.

The one-dimensional Hubbard model has been known to be exactly solvable
since the work of E.Lieb and F.Y.Wu of
1968 $^{\hbox{\scriptsize\cite{liebwu}}}$.
In their paper, a large set of eigenfunctions of the hamiltonian were found by
using the nested Bethe Ansatz $^{\hbox{\scriptsize\cite{yang}}}$. However,
the issue of whether this set of eigenfunctions is actually {\em complete}\
has not been considered until recently. In a recent paper
$^{\hbox{\scriptsize\cite{eks}}}$, we used the $SO(4)$ symmetry
of the Hubbard model (which has been explored in
${\hbox{{\cite{ya,yazh,yanew,eta,sofour}}}}$) to show that
the Bethe Ansatz is not complete (see below for some comments on this
result).

In this paper we will show that the $SO(4)$ structure can be used to
extend the set of Bethe Ansatz eigenstates to a complete set of eigenstates
of the one-dimensional Hubbard model. This result was first announced in
\cite{eks2}. Here we provide a detailed account of the derivations involved.

\vspace{6mm}

The Hubbard model describes electrons, which
can hop along a one-dimensional lattice and which interact with
coupling constant $U$ if
two of them occupy the same site.
The hamiltonian of the Hubbard model on a periodic one-dimensional
lattice of even, finite
length $L$ is given by (the small modifications in the potential term
as compared to
\cite{liebwu} are such that the $SO(4)$ symmetry becomes explicit,
see \cite{eta,ya,sofour,yazh})
\be
H = - \sum_{i=1}^{L} \sum_{\sigma=1,-1}
    (c^\dagger_{i,\si} c^{}_{i+1,\si} + c^\dagger_{i+1,\si} c^{}_{i,\si}) +
    U \sum_{i=1}^L(n_{i,1}-\half)(n_{i,-1}-\half) \,.
\label{hamil}
\ee
Here the $c_{i,\sigma}$ ($\sigma$ takes the values $\pm 1$) are
canonical Fermi operators on the lattice ($i=1,...,L$ labels the
lattice sites),
with anti-commutation relations given by $\{ c^\dagger_{i,\sigma} ,
c_{j,\tau} \} = \delta_{i,j} \delta_{\sigma,\tau}$.
They act in a Fock space with the pseudo vacuum $\mid \! 0 \rangle$
defined by $c_{i,\sigma} \mid \! 0 \rangle =0$. The operator $n_{i,\si}=
c^\dagger_{i,\si} c_{i,\si}$ is the number operator for electrons with
spin $\si$ on site $i$. $U$ is the coupling constant and can be either
positive (repulsive case) or negative (attractive case). For later
convenience we define
\be
u= {U\over 2i}\,.
\ee
The analysis by E.Lieb and F.Y.Wu in \cite{liebwu} resulted in a large
number of eigenstates of the hamiltonian, which are characterized by
momenta $k_i$ and rapidities $\La_\alpha$, where $i=1,2,\ldots,N_e$
and $\alpha=1,2,\ldots,M$ for an eigenstate with a total number of
$N$ spin-up and $M$ spin-down electrons. Our convention throughout the
paper will be
\hfill\break
\null\qquad $N$= number of spin-up electrons
\hfill\break
\null\qquad $M$= number of spin-down electrons
\hfill\break
\null\qquad $N_e=N+M$= total number of electrons\ .
\vskip .5cm
We will now discuss the Bethe wave functions
wave functions in the form as given by Woynarovich \cite{wor}, which
is equivalent to the form found by Lieb and Wu.
The nested Bethe Ansatz provides us with the following set of eigenstates
with M spins down and N spins up :
\bea
|\Psi_{_M,_N}\rangle &=& \sum_{1\leq x_k \leq L}
\psi_{_{-1,...,-1,1,...,1}}\xvec
\ \prod_{j=1}^M  c^\dagger_{x_j,-1}
\prod_{i=M+1}^{N_e} c^\dagger_{x_i,1} \ |0\rangle
\nonu
&=& \sum_{1\leq x_k \leq L}
    \psi_{_{\sigma_1,\sigma_2,...,\sigma_{N_e}}}\xvec
    \ \prod_{j=1}^{N_e}  c^\dagger_{x_j,\sigma_j}
    \ |0\rangle ,
\label{wfn}
\eea
where we have put $\sigma_1 = ... = \sigma_{M} =-1$,
$\sigma_{M+1}=...=\sigma_{N_e}=1$.

The Bethe Ansatz wave functions explicitly depend on the relative
ordering of the $x_{_i}$. We represent this dependence by a
permutation $Q$ of $N_e$ elements, which is such that
$1\leq x_{_{Q_1}}\leq x_{_{Q_2}}\le...\leq x_{_{Q_{N_e}}}\leq L$.
In the sector $Q$ the general Bethe wavefunction for $M$ spins down and $N$
spins up reads
\be
\psi_{_{\sigma_1,\sigma_2,...,\sigma_{N_e}}}\xvec = \hskip
-10pt\sum_{P\in S_{N_e}}\hskip -8pt sgn(Q)  \ sgn(P) \
e^{i\sum_{j=1}^{N_e}k_{P_j}x_{Q_j}}\ \varphi(\yvec|P) .
\label{wavefunc}
\ee
The $P$-summation extends over all permutations of $N_e$ elements and
$sgn(\Pi)$ is the sign of the permutation $\Pi$ ($\Pi = Q,P$).
The amplitudes $\varphi(\yvec|P)$ are of the form
\be
\varphi(\yvec|P) =\ \sum_{\pi\in S_{M}} \ A_{\pi} \prod_{l=1}^M
F_P(\Lambda_{\pi_l},y_l)
\ee
with
\bea
F_P(\Lambda_j,y) &=& \biggl(\ \prod_{i=1}^{y-1}\
{e^{(j)}_-(P_i)\over
e^{(j)}_+(P_i)} \ \biggr) \ {1\over e^{(j)}_+(P_y)}
\nonu
&=& \biggl(\ \prod_{i=1}^{y-1}\ {sin(k_{P_i})-\Lambda_j-{U\over 4i}
\over sin(k_{P_i})-\Lambda_j+{U\over 4i}}\biggr)
\left({1\over sin(k_{P_y})-\Lambda_j+{U\over 4i}}\right) \,,
\eea
where we defined
\be
e^{(j)}_\pm(i)= sin(k_i)-\Lambda_j\pm{u\over2}
\ee
and
\be
\strut{A_{\pi}\over\displaystyle\strut A_{(t,t+1)\pi}} =
{\strut{\Lambda_{\pi_{t+1}}-\Lambda_{\pi_{t}}-u}\over\displaystyle\strut
{\Lambda_{\pi_{t+1}}-\Lambda_{\pi_{t}}+u}}.
\label{Amp}
\ee
By $\pi = (\pi _1,\pi _2,..,\pi _t,\pi _{t+1},..\pi _M)$ we denote a
permutation of M elements (spin-down electrons) and ${\scriptstyle
(t,t+1)}\pi = (\pi _1,\pi _2,..,\pi _{t+1},\pi  _t,..\pi _M)$.
A solution of (\ref{Amp}) is given by
\be
A_\pi=\prod_{1\le l<k\le M}
\left(\frac{\Lambda_{\pi_l}-\Lambda_{\pi_k}+u}{\Lambda_{\pi_l}-\Lambda_{\pi_k}}
\right)\,.
\ee
The amplitudes $\varphi(\yvec|P)$ depend on $\sigma_1,...,\sigma_{N_e}$
and on $Q$ through the numbers $y_1,...,y_M$, which
are defined to be the positions of the down spins among the spins in the
series $\sigma_{Q_1},\sigma_{Q_2},...\sigma_{Q_{N_e}}$
in increasing order, {\em i.e.,}
\be
1\leq y_1< y_2< y_3< .....< y_M\leq N_e.
\ee
For example, for one spin down and one spin up ( and $\sigma_1=-1$,
$\sigma_2=1$ ) we have the two cases
$y=1$ (if the spin down is to the left, which holds in the
$Q = (id)$ sector) and $y=2$ (if the spin down is to
the right, which holds for $Q = (21)$).

As we already indicated in (\ref{wfn}), we will choose the notation such that
the $M$ down spins are at the positions $x_{_1},...,x_{_M}$, {\em i.e.,}
$\sigma_{_1}=...=\sigma_{_M}=-1$ and
$\sigma_{_{M+1}}=...=\sigma_{_{N_e}}=1$.

We see that all solutions are characterised by $N_e$ momenta
$\{k_j|{\scriptstyle j=1,...,N_e}\}$ of charged spinless excitations
(holons), and
$M$ rapidities $\{\Lambda_k|{\scriptstyle k=1,...M}\}$
of spin waves (spinons).

Imposing periodic boundary conditions on the Bethe Ansatz wavefunctions
leads to the following equations for the parameters $k_i$ and $\La_\alpha$
\bea
e^{ik_j L} = \prod_{\alpha=1}^{M}
  \frac{\sin(k_j) - \La_\alpha - \frac{U}{4i}}
       {\sin(k_j) - \La_\alpha + \frac{U}{4i}},
  \hspace{1cm} j=1,2,\ldots N_e \,,
\nonumber \\[4mm]
\prod_{j=1}^{N_e}
  \frac{\sin(k_j) - \La_\alpha - \frac{U}{4i}}
       {\sin(k_j) - \La_\alpha + \frac{U}{4i}}
  = - \prod_{\beta=1}^{M}
  \frac{\La_\beta - \La_\alpha - \frac{U}{2i}}
       {\La_\beta - \La_\alpha + \frac{U}{2i}},
  \hspace{1cm} \beta = 1,2,\ldots M \,.
\label{pbc}
\eea

Energy and momentum, {\em i.e.,} the eigenvalues of the
hamiltonian (\ref{hamil}) and the logarithm of the translation operator,
of the system in a state corresponding to a
solution of (\ref{pbc}) are
\bea
E_{_{N_e}} &=& -2 \sum_{i=1}^{N_e} cos(k_i)+{U\over 2}\
\left[{L\over 2}-N_e\right],
\nonu
P &=& \sum_{i=1}^{N_e} k_i .
\eea
The second term in the expression for the energy in due to the shift of
$n_{j,\sigma}$ by ${1\over 2}$ in (\ref{hamil}).

Because of the antisymmetry of the product over $c^\dagger$'s under
interchange of any two of them, the wave functions
$\psi_{\sigma_1...\sigma_{N_e}}(x_{_1}...x_{_{N_e}}) $
can be (and have been) chosen to be completely antisymmetric
under the simultaneous exchange $x_k \longleftrightarrow x_j$ and $\sigma_k
\longleftrightarrow \sigma_j$, {\em i.e.,}
\be
\psi_{\sigma_1..\sigma_j..\sigma_k..\sigma_{N_e}}
(x_{_1}...x_{_j}...x_{_k}...x_{_{N_e}})
= -
\psi_{\sigma_1..\sigma_k..\sigma_j..\sigma_{N_e}}
(x_{_1}...x_{_k}...x_{_j}...
x_{_{N_e}}) \,.
\ee

We now define `regular' Bethe Ansatz
states (for finite $L$), to be denoted by $\mid\! \psi_{M,N}\rangle$, by the
properties that $N-M \geq 0$ (non-negative third component of the
spin), $N_e \leq L$
(less than or equal to half filling), and that all $\Lambda_\alpha$ and
all $k_j$ are
finite. Bethe Ansatz states with $N-M<0$ and/or $N_e > L$ can be obtained
from the regular Bethe Ansatz states by using simple symmetry
operations, {\sl i.e.,} reflection of the third component of the spin
and particle/hole correspondence,
which commute with the hamiltonian $^{\hbox{\scriptsize\cite{suth}}}$.
The model is invariant under spin-rotations, with the corresponding
$SU(2)$ generators given by
\be
\zeta = \sum_{i=1}^L c^\dagger_{i,1} c^{}_{i,-1}, \hspace{1cm}
\zeta^\dagger = (\zeta)^\dagger, \hspace{1cm}
\zeta_z = \half \sum_{i=1}^L (n_{i,-1}-n_{i,1}).
\label{so1}
\ee
(Note that $\zeta_z$ equals {\em minus} the third component of
the total spin.)
For even $L$ the model has a second $SU(2)$ invariance, generated
by $^{\hbox{\scriptsize\cite{yazh}}}$
\be
\eta = \sum_{i=1}^L (-1)^i c_{i,1} c_{i,-1}, \hspace{5mm}
\eta^\dagger
   = \sum_{i=1}^L (-1)^i c^\dagger_{i,-1} c^\dagger_{i,1}, \hspace{5mm}
\eta_z = \half \sum_{i=1}^L (n_{i,-1}+n_{i,1})-\frac{L}{2}.
\label{so2}
\ee
The raising operator $\eta^\dagger$ of this second $SU(2)$ creates
a pair of two opposite-spin electrons on the same site, with momentum $\pi$.
Combining the two $SU(2)$'s, which commute with the hamiltonian
and with one another, leads to an $SO(4)$ invariance of the one-dimensional
Hubbard model for even lattice
lengths$^{\hbox{\scriptsize\cite{sofour,yazh}}}$.
For a discussion of the
theoretical and possible experimental
consequences of the existence of this symmetry, which also exists in the
Hubbard model in two or three dimensions, we refer to the papers
\cite{yazh,yanew,sofour,zhang}.

In a previous paper $^{\hbox{\scriptsize\cite{eks}}}$, we established the
following remarkable property of the regular Bethe Ansatz eigenstates of the
hamiltonian: they are all {\em lowest weight states} of the $SO(4)$ algebra
(\ref{so1}), (\ref{so2}), {\em i.e.},
\be
    \eta \mid \psi_{M,N}\rangle \, = 0 \,, \hspace{1cm}
    \zeta \mid \psi_{M,N}\rangle\, = 0.
\ee
This shows that acting with the raising operators $\eta^\dagger$ and
$\zeta^\dagger$ on $\mid \psi_{M,N}\rangle$ leads to new eigenstates of the
hamiltonian, which are not in the regular Bethe Ansatz.
In this way, every regular Bethe Ansatz state $\mid \psi_{M,N}\rangle$ is the
lowest weight state in a multiplet of states, which form a representation
of $SO(4)$. Since
\be
    \eta_z \mid \psi_{M,N}\rangle \, = \half (N_e-L) \mid \psi_{M,N}\rangle \,,
    \hspace{6mm}
    \zeta_z \mid \psi_{M,N}\rangle \, = \half (M-N) \mid \psi_{M,N}\rangle \,,
\ee
a state $\mid\! \psi_{M,N}\rangle$ has spin $\eta=
\half (L-N_e)$ with respect to the $\eta$-pairing $SU(2)$ algebra
and spin $\zeta = \half (N-M)$ with respect to the $\zeta$ $SU(2)$
algebra. The dimension of the corresponding $SO(4)$ multiplet is
therefore given by
\be
     {\rm dim}_{M,N} = (2\, \eta+1)(2\, \zeta+1) = (L-N_e+1)(N-M+1) \,.
\label{dim}
\ee
The states in this multiplet are of the form
\be
|\psi_{M,N}^{\alpha,\beta}\rangle=(\eta^\dagger)^\alpha (\zeta^\dagger)^\beta
\mid \psi_{M,N}\rangle\,.
\ee
By symmetry, all the states that
are highest or lowest weight states with respect to one of the
$SU(2)$ algebras are again given by the Bethe Ansatz (although
in general they are outside the {\em regular}\ Bethe Ansatz). All other
states are {\em not} given by the Bethe Ansatz, which shows that for
this model the Bethe Ansatz is not complete. The simplest example of
a state that is outside the Bethe Ansatz is $\eta^\dagger \mid 0\rangle$,
which describes a single $\eta$-pair of momentum $\pi$. The fact that
this state is outside the Bethe Ansatz was explicitly confirmed in
\cite{eks}.

It is the main purpose of this paper to show that,
if one counts the number of eigenstates
that are related to the regular Bethe Ansatz states by the $SO(4)$ symmetry,
one finds $4^L$, which is precisely the correct dimension of the Hilbert
space of the model \footnote{There are 4 possible electron
configurations per lattice site (spin up, spin down, both spin up and
spin down, and empty site), thus the corresponding direct product
Hilbert space is $4^L$ dimensional.}. Thus we will conclude that the
Bethe Ansatz together with the $SO(4)$ structure leads to a complete
set of eigenstates of the one-dimensional Hubbard model.

The paper is organized as follows. In section 2 we discuss in some detail
the nature of the solutions of the Bethe equations for the Hubbard model.
In Appendix A we show that the so-called $\La$ and $k-\La$ strings
give wave functions that describe bound states.
In section 3 we will then count eigenstates and prove completeness.
An explicit construction of the $\half (L+2)(L-1)$ solutions of the
Bethe equations in the sector with one spin-up and one spin-down electron
(as opposed to the somewhat indirect construction used in the general
proof) is presented in Appendix B.

\section{Solutions of the Bethe equations for the Hubbard model}
\setcounter{equation}{0}

Let us focus on the Bethe Equations (\ref{pbc}), which express the
fact that the Bethe Ansatz wavefunctions (\ref{wavefunc}) satisfy
periodic boundary conditions.

Counting regular Bethe Ansatz states means counting inequivalent solutions
of the equations (\ref{pbc}) while taking into account the `regularity
conditions' $N-M \geq 0$ and $N_e \leq L$. Following M.Takahashi
$^{\hbox{\scriptsize\cite{takahub}}}$, we will first distinguish different
types of solutions $\{k_i,\La_\alpha \}$ of (\ref{pbc}).
The idea is that for a solution $\{ k_i,\La_\alpha \}$, the set of all
the $k_i$'s and $\La_\alpha$'s can be split into (three) different kinds
of subsets (`strings'), which are
\begin{enumerate}
\item
  a single real momentum $k_i$
\item
  $m$ $\La_\alpha$'s combine into a string-type configuration
  (`$\La$-strings'); this includes the case $m=1$, which is just a
  single real $\La_\alpha$
\footnote{these correspond to bound states of spin waves (magnons)}
\item
  $2m$ $k_i$'s and $m$ $\La_\alpha$'s combine into a different
   string-type configuration (`$k$-$\La$-strings')
\footnote{the case $m=1$ describes a `Cooper pair' of electrons}.
\end{enumerate}
For large lattices ($1<<L$), almost all the string configurations
are close to `idealized' string-solutions where both the $k$'s and the
$\La$'s are assigned imaginary parts according to a `equal-spacing'
prescription $^{\hbox{\scriptsize\cite{takahub}}}$. For a $\La$-string
of length $m$ the rapidities involved are
\be
\La_{\alpha}^{m,j} = \La_{\alpha}^m - \frac{1}{2} (m+1-2j) u
\qquad \La_{\alpha}^m {\ \rm real}\qquad j=1,2, \ldots, m \,.
\label{lambdastr}
\ee
The $k$'s and $\La$'s involved in a $k$-$\La$-string are
\bea
k_\alpha^1 &=& \pi - \sin^{-1}(\La_\alpha^{\prime \, m} - \frac{1}{2} m u)
\nonu
k_\alpha^2 &=& \sin^{-1}(\La_\alpha^{\prime \, m} - \frac{1}{2} (m-2) u)
\nonu
k_\alpha^3 &=& \pi - k_\alpha^2
\nonu
k_\alpha^4 &=& \sin^{-1}(\La_\alpha^{\prime \, m} - \frac{1}{2} (m-4) u)
\nonu
k_\alpha^5 &=& \pi - k_\alpha^4
\nonu
\cdots
\nonu
k_\alpha^{2m-2} &=& \sin^{-1}(\La_\alpha^{\prime \, m} + \frac{1}{2} (m-2) u)
\nonu
k_\alpha^{2m-1} &=& \pi - k_\alpha^{2m-2}
\nonu
k_\alpha^{2m} &=&
  \pi - \sin^{-1}(\La_\alpha^{\prime \, m} + \frac{1}{2} m u) \,.
\eea
and
\be
\La_{\alpha}^{\prime \, m,j} = \La_{\alpha}^{\prime \, m}
  - \frac{1}{2} (m+1-2j) u \ , \quad \La_{\alpha}^{\prime\, m}
{\ \rm real}\qquad j=1,2, \ldots, m \,.
\label{klstr}
\ee
(\ref{lambdastr}) - (\ref{klstr}) are valid up to exponential corrections
of order ${\cal O}(\exp{-\delta L})$, where $\delta$ is real and
positive (and depends on the specific string under consideration).
In Appendix A we discuss the wave-functions corresponding to some
of the the string configurations 2. and 3. and show that they
correspond to bound states.

Let us now consider a solution that splits into $M_m$
$\La$-strings of length $m$, $M^\prime_n$ $k$-$\La$-strings
of length $n$ (containing $2n$ $k_i$'s and $n$ $\La_\alpha$'s) and $M_e$
additional single $k_i$'s. Clearly, we have
\be
N_e = M_e + 2\, \sum_{m=1}^{\infty} m \, M^\prime_m \,,
\hspace{6mm}
M = \sum_{m=1}^{\infty} m (M_m + M_m^\prime) \,.
\label{cons}
\ee
How many solutions of this type exist?

The idea is that each of the strings in a solution can be characterised
by the position of its center (a real number), which we denote as in
(\ref{lambdastr}) by $\La_\alpha^m$,
$\alpha=1,2,\ldots M_m$, for the length-$m$ $\La$-strings,
by $\La_\alpha^{\prime m}$, $\alpha=1,2,\ldots M^\prime_m$ for the
length-$m$ $k$-$\La$-strings (as in (\ref{klstr}) ) and which is simply
equal to $k_j$ for the unpaired momenta $k_j$, $j=1,2,\ldots,M_e$.
Because of the periodic boundary conditions, these parameters will
have to be chosen from a discrete set.

Following \cite{takahub}, we now write the following equations for the
parameters $k_j$, $\La_\alpha^m$ and $\La_\alpha^{\prime m}$. They
follow from (\ref{pbc}) and the form of the `idealized'
string-solutions which we discussed above (we write $M^\prime =
\sum_{m=1}^\infty m M_m^\prime$)
\bea
&& k_j L =
   2 \pi I_j
 - \sum_{n=1}^\infty \sum_{\alpha =1}^{M_n}
   \theta \left( \frac{\sin k_j - \Lambda_\alpha^n}{n U} \right)
 - \sum_{n=1}^\infty \sum_{\alpha =1}^{M^\prime_n}
   \theta \left( \frac{\sin k_j - \Lambda_\alpha^{\prime n}}
                      {n U} \right) \,,
\nonumber \\[4mm]
&& \sum_{j=1}^{N_e-2M^\prime}
   \theta \left( \frac{\Lambda_\alpha^n - \sin k_j}
                      {n U} \right)
 = 2 \pi  J_\alpha^n
   + \sum_{m=1}^\infty\sum_{\beta =1}^{M_m} \Theta_{nm} \left(
     \frac{\Lambda_\alpha^n - \Lambda_\beta^m} {U} \right) \,,
\nonumber \\[4mm]
&& L \left[ sin^{-1} \left( \Lambda_\alpha^{\prime n} + i n \frac{U}{4} \right)
  + sin^{-1} \left( \Lambda_\alpha^{\prime n} - i n \frac{U}{4} \right) \right]
\nonu
&& \hspace{1cm} =
   2 \pi J_\alpha^{\prime n}
 + \sum_{j=1}^{N_e-2M^\prime}
   \theta \left( \frac{\Lambda_\alpha^{\prime n} - \sin k_j}
                      {n U} \right)
 + \sum_{m=1}^\infty\sum_{\beta =1}^{M_m^\prime} \Theta_{nm} \left(
     \frac{\Lambda_\alpha^{\prime n} - \Lambda_\beta^{\prime m}}
                      {U} \right) \,,
\label{logpbc}
\eea
where
\bea
\theta(x) &=& 2 \tan^{-1} (4x)
\nonumber \\[4mm]
\Theta_{nm}(x) &=& \!\!
\left\{ \begin{array}{l}
      \theta \left( \frac{4\, x}{\mid n-m \mid} \right)
      + 2 \theta \left( \frac{4\, x}{\mid n-m \mid +2} \right)
      + \ldots
      + 2 \theta \left( \frac{4\, x}{n+m-2} \right)
      + \theta \left( \frac{4\, x}{n+m} \right)
\\
      \str \hspace{9.1cm} {\rm for} \;\; n \neq m
\\
     2 \theta \left( 2 \, x\over 1 \right) + \ldots
     +  2 \theta \left( \frac{2\, x}{n-1} \right)
     +  \theta \left( \frac{2\, x}{n} \right)
     \hspace{3cm} {\rm for} \;\; n = m
\end{array} \right.
\nonu &&
\eea
The $I_j$, $J_\alpha^n$ and $J_\alpha^{\prime n}$ are integer or
half-odd-integer according to the following prescriptions: $I_j$ is integer
(half-odd-integer) if $\sum_m (M_m+M^\prime_m)$ is even (odd); the $J_\alpha^n$
are integer (half-odd-integer) if $(N_e-M_n)$ is odd (even); the
$J_\alpha^{\prime n}$ are integer (half-odd-integer) if $(L-(N_e-M_n^\prime))$
is odd (even). According to \cite{takahub}, we have the following inequalities
\bea
\mid J_\alpha^n \mid &\leq& \half (N_e - 2 M^\prime
     - \sum_{m=1}^\infty t_{nm} M_m -1) \,,
\nonu
\mid J_\alpha^{\prime n} \mid &\leq& \half (L - N_e + 2 M^\prime
     - \sum_{m=1}^\infty t_{nm} M_m^\prime -1) \,,
\nonu
0<I_j&\le& L\,,
\label{ineq}
\eea
where $t_{nm} = 2 \,{\rm Min}(n,m) - \de_{nm}$.

We will now make the standard assumption that, in order to enumerate
the different solutions of the system (\ref{logpbc}), it is sufficient
to enumerate all possible sets of {\sl non-repeating} (half-odd)\-integers
$I_j$, $J_\alpha^n$ and $J_\alpha^{\prime n}$, satisfying (\ref{ineq}).

[This assumption mimics the similar assumption which is usually made
for the spin-${1\over 2}$ Heisenberg XXX model \cite{fadtak,takaxxx}.
It is known, however\footnote{This fact was actually already noticed
in the original paper by Bethe, \cite{bethe}.},
that the actual distribution of the different types of solutions can be
different from the one implied by this counting. We carefully studied
this phenomenon in a recent paper \cite{eks3}, where we give a detailed
discussion of the two-magnon sector of the XXX model. We explicitly show
that the deviations from the above assumption can be viewed as a
`redistribution phenomenon', which does not affect the total number of
Bethe Ansatz states. For the Hubbard model we have a similar situation,
which gets complicated further by the fact that there is a free coupling
constant $U$ in the model. In Appendix B we analyze in detail the $N=1$,
$M=1$ sector of the Hubbard model. Although we do find $U$-dependent
redistributions among different types of solutions, we find agreement
with the predictions based on the `ideal' string assumption for
the total number of states in this sector, which is $\half (L-1)(L+2)$ for
the $L$-site model.]

\hskip -3pt From (\ref{ineq}) we read off that the numbers of allowed values
for the (half-odd-)integers corresponding to each of the fundamental strings
are
\begin{enumerate}
\item
 $L$ for a free $k_i$
\item
$N_e-2 M^\prime - \sum_{m=1}^{\infty} t_{nm} M_m$ for a $\La$-string of
length $n$
\item
$L- N_e + 2 M^\prime - \sum_{m=1}^{\infty} t_{nm} M^\prime_m$
for a $k$-$\La$-string of length $n$.
\end{enumerate}
The total number of ways to choose the (half-odd-)integers in a solution
with multiplicities $M_e$, $M_m$ and $M_m^\prime$ is therefore given
by (remember that the integers are assumed to be non-repeating)
\bea
&& \hskip -2pt n(M_e, \{ M_m \} , \{ M_m^\prime \})=
 \left(\begin{array}{c} \str L \\ \str M_e
        \end{array} \right)
\prod_{n=1}^{\infty} \left( \begin{array}{c}
   \str N_e - 2 M^\prime
            - \sum_{m=1}^\infty t_{nm} M_m
   \\ \str M_n\end{array}\right)\times
\nonu && \hspace{1.5cm} \times
\prod_{n=1}^{\infty} \left( \begin{array}{c}
   \str L- N_e + 2 M^\prime
   - \sum_{m=1}^\infty t_{nm} M^\prime_m
   \\ \str M_n^\prime
                           \end{array} \right) \,.
\label{number}
\eea

The total number of solutions of (\ref{pbc}) with given numbers $N$ and
$M$ is now obtained by summing
$n(M_e, \{ M_m \} , \{ M_m^\prime \})$ over all the
$M_e$, $M_m$ and $M_m^\prime$, under the constraints (\ref{cons}).

Every solution to
(\ref{pbc}) gives us a regular Bethe Ansatz state, which comes
with an entire multiplet of eigenstates of the hamiltonian, the
dimension ${\rm dim}_{M,N}$ of which is given in (\ref{dim}).
The full number of eigenstates that are obtained from the
Bethe Ansatz {\em and} the $SO(4)$ symmetry is therefore given by
\bea
\lefteqn{\# \; ({\rm eigenstates})=}
\nonu
&& \begin{array}{cccccc}
   \displaystyle{ \sum_{M\ge 0} \sum_{N\ge 0} } &
   \left[ \Str \right. &
   \displaystyle{ \sum_{M_e=0}^\infty \sum_{M_m=0}^\infty
        \sum_{M_m^\prime=0}^\infty } &
   n(M_e, \{ M_m \} , \{ M_m^\prime \}) &
   \left. \Str \right] &
   {\rm dim}_{M,N} \,. \\
\strut   \scriptstyle{N-M \geq 0} & &
   \scriptstyle{N+M = M_e + 2\, \sum_{m=1}^\infty m M^\prime_m} & & & \\
   \scriptstyle{N+M \leq L} & &
   \scriptstyle{N-2M = M_e-2\sum_{m=1}^\infty m M_m} & & &
   \end{array}
\label{sum}
\eea

The counting of the eigenstates that are obtained from the $SO(4)$
extended nested Bethe Ansatz has thus been reduced to a purely algebraic
problem, which we will solve in the next section.

\section{Counting eigenstates}
\setcounter{equation}{0}

In this section we will prove that for general even $L$ the sum
in (\ref{sum}) equals $4^L$. This will prove completeness.
Before we come to that, we show the
examples of the 2-site and 4-site models. The 2-site model
($L=2$) was discussed in \cite{eks}, where we presented the
explicit form of a complete set of $4^2=16$ eigenstates of the
hamiltonian. In Table~1\ we show how the counting presented in
section 3 works out in this case.

\begin{table}[h]
\begin{center}
\begin{tabular}{|ccc|cc|c|c|r|}
\hline
$\STr$ $M_e$ & $M_1$ & $M_1^\prime$ & $M$ & $N$
     & $n$ & dim$_{M,N}$ & \#(states) \\
\hline
$\str$ 0 & 0 & 0 &   0 & 0 &   1 & 3 & 3 \\
$\str$ 1 & 0 & 0 &   0 & 1 &   2 & 4 & 8 \\
$\str$ 2 & 0 & 0 &   0 & 2 &   1 & 3 & 3 \\
$\str$ 2 & 1 & 0 &   1 & 1 &   1 & 1 & 1 \\
$\str$ 0 & 0 & 1 &   1 & 1 &   1 & 1 & 1 \\
\hline
$\str$   &   &   &     &   &     &   & 16 \\
\hline
\end{tabular}
\caption{ $L=2$. $n$ denotes the number of regular
   Bethe Ansatz states of a given type. There are a total number of
   $16=4^2$ eigenstates of the hamiltonian.}
\end{center}
\end{table}

\begin{table}
\begin{center}
\begin{tabular}{|ccccc|cc|c|c|r|}
\hline
$\STr$ $M_e$ & $M_1$ & $M_2$ & $M_1^\prime$ & $M_2^\prime$ &
     $M$ & $N$ & $n$ & dim$_{M,N}$ & \#(states) \\
\hline
$\str$ 0 & 0 & 0 & 0 & 0 &   0 & 0 &   1 & 5 & 5  \\
$\str$ 1 & 0 & 0 & 0 & 0 &   0 & 1 &   4 & 8 & 32 \\
$\str$ 2 & 0 & 0 & 0 & 0 &   0 & 2 &   6 & 9 & 54 \\
$\str$ 3 & 0 & 0 & 0 & 0 &   0 & 3 &   4 & 8 & 32 \\
$\str$ 4 & 0 & 0 & 0 & 0 &   0 & 4 &   1 & 5 & 5  \\
$\str$ 2 & 1 & 0 & 0 & 0 &   1 & 1 &   6 & 3 & 18 \\
$\str$ 0 & 0 & 0 & 1 & 0 &   1 & 1 &   3 & 3 & 9  \\
$\str$ 3 & 1 & 0 & 0 & 0 &   1 & 2 &   8 & 4 & 32 \\
$\str$ 1 & 0 & 0 & 1 & 0 &   1 & 2 &   8 & 4 & 32 \\
$\str$ 4 & 2 & 0 & 0 & 0 &   2 & 2 &   1 & 1 & 1  \\
$\str$ 4 & 0 & 1 & 0 & 0 &   2 & 2 &   1 & 1 & 1  \\
$\str$ 2 & 1 & 0 & 1 & 0 &   2 & 2 &   6 & 1 & 6  \\
$\str$ 0 & 0 & 0 & 2 & 0 &   2 & 2 &   1 & 1 & 1  \\
$\str$ 0 & 0 & 0 & 0 & 1 &   2 & 2 &   1 & 1 & 1  \\
$\str$ 4 & 1 & 0 & 0 & 0 &   1 & 3 &   3 & 3 & 9  \\
$\str$ 2 & 0 & 0 & 1 & 0 &   1 & 3 &   6 & 3 & 18 \\
\hline
$\str$   &   &   &   &   &     &   &     &   & 256 \\
\hline
\end{tabular}
\caption{$L=4$. There are 60 regular Bethe Ansatz states, which,
when weighted with the correct $SO(4)$ multiplicities, give a total of
$256=4^4$ eigenstates of the hamiltonian.}
\end{center}
\end{table}
The total number of 16 states
splits into 2 singlets, 2 triplets and 2 quadruplets of $SO(4)$.
The ground state is the singlet with $M_1 = 1$ for the case $U>0$ and
the singlet with $M_1^\prime =1$ for the case $U<0$.
In both cases it is a
bound state of one spin up and one spin down electron with energy
$E_0 = - \sqrt{U^2/4 + 16}$. The counting for the 4-site model
($L=4$) is presented in Table~2, where we show how the total
number of $4^4=256$ is obtained. (Notice that the total number
of regular Bethe Ansatz states is only 60.)

We now turn to the proof that for general (even) $L$ the
sum in (\ref{sum}) equals $4^L$. We will split this proof into
two steps as follows. In the first step we will prove the following
two identities
\be
\begin{array}{c}
         \str \\
         \displaystyle{ \sum_{M_1,M_2,\ldots =0}^\infty} \\
         \str \scriptstyle{\sum_{m=1}^\infty m M_m = M}
\end{array}
\prod_{n=1}^\infty
   \left( \begin{array}{c} \Str N-\sum_m t_{nm} M_m \\
                           \Str M_n    \end{array}  \right) =
  \left( \begin{array}{c} N \\ M \end{array} \right)
  - \left( \begin{array}{c} N \\ M-1 \end{array} \right)
\label{xxx1}
\ee
and
\be
  \sum_{M=0}^{[N/2]}
  \left( \left( \begin{array}{c} N \\ M \end{array} \right)
         - \left( \begin{array}{c} N \\ M-1 \end{array} \right)
  \right) (N - 2 \, M +1) = 2^N \,.
\label{xxx2}
\ee
For later convenience we define
\be
P_n= N-\sum_{m=1}^\infty t_{nm} M_m\,,
\label{pn}
\ee
and
\be
n(\{ M_m \})=
\prod_{n=1}^\infty
   \left( \begin{array}{c} \Str P_n \\
                           \Str M_n    \end{array}  \right)\,,
\ee
where $t_{nm} = 2 \,{\rm Min}(n,m) - \de_{nm}$ as before.
In the second step we will then use identities (\ref{xxx1}) and (\ref{xxx2})
to perform the summation in (\ref{sum}).

The auxiliary identities (\ref{xxx1}) and (\ref{xxx2}) have a natural
interpretation in the context of the spin-$1/2$ Heisenberg XXX
model \cite{takaxxx}.
The equation (\ref{xxx1}) gives the total number of regular Bethe Ansatz
states (defined by $M \leq [N/2]$)
with $M$ overturned spins in the XXX model on a lattice of length
$N$. The second formula shows that the total number
of states obtained by combining the regular Bethe Ansatz with the $SU(2)$
structure equals $2^N$, which is the dimension of the Hilbert space of the
XXX model. These relations thus establish the completeness
of the $SU(2)$ extended Bethe Ansatz for the XXX model.

The fact that identities that have their origin in the XXX model play
a role here should not come as a surprise. Indeed, our method of
solution of the Hubbard model is the {\it nested} Bethe Ansatz.
The solutions to the Bethe Ansatz are specified by two sets $\{k_j\}$ and
$\{\Lambda_\alpha\}$ of spectral parameters.
The $k_j$'s are momenta of charge density waves, whereas the
$\Lambda_\alpha$'s, which describe the `nesting' of the Bethe Ansatz, are
rapidities of spin density waves of the type encountered in the
Heisenberg XXX model.
This should make clear that the second stage of the nested Bethe Ansatz
for the Hubbard model is really a spin-problem, which is very similar
to the Bethe Ansatz analysis of the Heisenberg XXX model.
Our two-step procedure for performing the summation is natural
from the point of view of the nesting: in the first step we sum
over the spin degrees of freedom, and in the second step we then
sum over the charge degrees of freedom as well.

\vspace{5mm}

\noindent {\sc STEP 1.}

\vspace{5mm}

Let us now explain how the equation (\ref{xxx1}) can be derived.
In the first step, one simply solves for $M_1=M-\sum_{m=2}^\infty m M_m$
and substitutes this back into the left hand side of (\ref{xxx1}).
Using this value for $M_1$, the quantities $P_n$ reduce to
\bea
&& P_1 = N-\sum_{m=1}^\infty t_{1m} M_m = N-M + \sum_{m=3}^\infty (m-2) M_m \,,
\nonu
&& P_n = N-\sum_{m=1}^\infty t_{nm} M_m = N -2M +M_n +2
\sum_{m=n+1}^\infty (m-n) M_m\,.
\eea
Let us now consider the summation over $M_2$. Although our
summand in the left hand side of (\ref{xxx1}) has the form of
an infinite product, only two of the factors contain the variable
$M_2$. Singling these out, one finds that the summation over $M_2$
is as follows
\bea
\Omega_2 = \sum_{M_2=0}^\infty &&
   \left( \begin{array}{c}
       \Str N-2M+M_2+2 \sum_{m=3}^\infty (m-2) M_m \\ \Str M_2
   \end{array} \right) \times
\nonu
   && \times \left( \begin{array}{c}
      \Str  N -M + \sum_{m=3}^\infty (m-2) M_m \\
      \Str M- \sum_{m=2}^\infty m M_m
   \end{array} \right) .
\label{omII}
\eea
In order to perform this summation
we will make use of the identity
\be
\sum_{\alpha=0}^\infty \left( \begin{array}{c} B+\alpha \\ \alpha
    \end{array} \right) x^\alpha = (1-x)^{-1-B} \,,
\label{binom}
\ee
which can easily be proved by induction.
As a simple consequence, we have
\be
(1-x^2)^{-1-\omega}(1+x)^\eta \mid_{x^A}
= \sum_{\alpha=0}^\infty \left( \begin{array}{c} \omega + \alpha
   \\ \alpha \end{array} \right) \left( \begin{array}{c}
   \eta \\ A- 2\alpha \end{array} \right) \,,
\label{xeqn}
\ee
where the notation $ \mid_{x^A}$ in the left hand side means that we
single out the coefficient of the power $x^A$.
The right hand side of (\ref{omII}) is of the same form as (\ref{xeqn})
and we find
\bea
\Omega_2 &=& (1-x^2)^{-1-[N-2M+2(M_3+ 2 M_4 +\ldots)]}
  (1+x)^{N-M+ (M_3+2 M_4+ \ldots)} \mid_{x^{M-3M_3-4M_4-\ldots}}
\nonu
&=& (1+x)^{N-M}(1-x^2)^{-N+2M-1} \, \prod_{n=3}^\infty
  ( {\cal Z}_n^{(0)} )^{M_n} \mid_{x^M}
\nonu
&=& {1\over 2 \pi i}\oint\frac{dx}{x^{M+1}}
(1+x)^{N-M}(1-x^2)^{-N+2M-1}\prod_{n=3}^{\infty}({\cal Z}_n^{(0)})^{M_n}\,,
\label{o2}
\eea
where
\be
    {\cal Z}_n^{(0)} = \frac{x^n}{(1-x)^{2(n-2)}(1+x)^{n-2}} \,.
\label{znnull}
\ee
In the last line of (\ref{o2}) we extracted the coefficient at $x^M$
by performing a contour integral around the origin $x=0$.
After performing the $M_2$ summation (\ref{xxx1}) now reads
\be
\begin{array}{c}
         \str \\
         \displaystyle{\sum_{M_1,M_2,\ldots =0}^\infty} \\
         \str \scriptstyle{\sum_{m=1}^\infty m M_m = M}
\end{array} n(\{ M_m \})
=\frac{1}{2 \pi i} \oint \frac{dx}{x^{M+1}} A(x)\,,
\label{avonx0}
\ee
where
\bea
A(x) &=& (1+x)^{N-M}(1-x^2)^{-1-N+2M}\quad\times
\nonu
\times
\sum_{M_3,M_4,\ldots =0}^\infty \prod_{n=3}^\infty &&\hskip -30pt
  \left( \begin{array}{c} \Str N-2M +M_n +2 \sum_{m=n+1}^\infty (m-n)M_m
  \\ \Str M_n \end{array} \right)
\prod_{l=3}^\infty( {\cal Z}_l^{(0)})^{M_l}\,.
\label{avonx1}
\eea

The summation over $M_3$ is given by
\bea
\Omega_3 &=& \sum_{M_3=0}^\infty
    \left( \begin{array}{c} \Str N-2M+M_3+2 \sum_{n=4}^\infty (n-3)M_n
       \\ \Str M_3 \end{array} \right) ( {\cal Z}_3^{(0)} )^{M_3}
\nonu
&=& \left( 1 - {\cal Z}_3^{(0)} \right)^{-1-N+2M-2\sum_{n=4}^\infty
    (n-3) M_n} \,.
\eea
At this point, the full expression for $A(x)$ has been reduced to
\bea
\lefteqn{A(x) = (1+x)^{N-M}(1-x^2)^{-1-N+2M}
   \left( 1 - {\cal Z}_3^{(0)} \right)^{-1-N+2M} \times}
\nonu
&& \times \sum_{M_4,M_5,\ldots =0}^\infty \prod_{n=4}^\infty
  \left( \begin{array}{c} \Str N-2M +M_n +2 \sum_{m=n+1}^\infty (m-n)M_m
  \\ \Str M_n \end{array} \right)
  \prod_{l=4}^\infty ( {\cal Z}_l^{(1)} )^{M_l} \,,
\nonu
&&
\eea
where ${\cal Z}_n^{(1)}$ are defined through
\be
{\cal Z}_n^{(1)} = \frac{ {\cal Z}_n^{(0)} }
  { \left( 1 - {\cal Z}_3^{(0)} \right)^{2(n-3)} }   \,.
\ee

{}From the above it is now clear, that the sum with respect to $M_4$ and
all $M_n$ with $n>4$ has the same structure as the sum with respect to
$M_3$. Thus the final result, after performing all
summations, will look like
\be
A(x) = (1+x)^{N-M} F(x)^{-1-N+2M} \,,
\label{avonx2}
\ee
where
\be
F(x) = (1-x^2) \prod_{m=3}^\infty  \left( 1 - {\cal Z}_m^{(m-3)} \right)
\ee
and we have the iteration formula
\be
{\cal Z}_n^{(m)} = \frac{ {\cal Z}_n^{(m-1)} }
  { \left( 1 - {\cal Z}_{m+2}^{(m-1)} \right)^{2(n-m-2)} }   \,.
\label{iter}
\ee

Our task is now to find a closed expression for $F(x)$ by exploiting this
relation. We define
\be
U_2=x^{-2}\,,
\quad  U_m = \frac{1}{ {\cal Z}_m^{(m-3)} }\,, \quad m\geq 3  \,,
\ee
so that $F(x)$ can be written as
\be
F(x) = \prod_{m=2}^\infty  \left( 1 - \frac{1}{U_m} \right)  \,.
\label{fx}
\ee
We now claim that the functions $U_m(x)$ satisfy the following
recursion relation, to be denoted by $RR\ I_p$
\be
RR\ I_p: \qquad (U_{p+3}-1)^2 = U_{p+4} U_{p+2} \,, \quad p\geq 0 \,.
\label{recur1}
\ee
Together with the initial conditions
\be
U_2 = x^{-2}, \quad U_3 = \frac{(1-x)^2(1+x)}{x^3}
\label{init}
\ee
these relations completely fix the functions $U_m(x)$ and thereby the
function $F(x)$.

In order to prove the recursion relation (\ref{recur1}), we first
give a second recursion relation, which involves some of the other
${\cal Z}$'s and which we shall denote by $RR\ II_p$
\be
RR\ II_p: \qquad \frac{ {\cal Z}_{n+1}^{(p)}}{ {\cal Z}_{n}^{(p)}} =
\frac{ U_{p+2} }{ U_{p+3} } \,, \quad p\geq 0, \;\;  n\geq p+3 \,.
\label{recur2}
\ee
Let us now show that the validity of both recursion relations
can be proved by induction. We start at the point where we
have ${\cal Z}_{n}^{(0)}$, which is defined by (\ref{znnull}), and
$U_2$ and $U_3$ as above in (\ref{init}).
One easily checks that $RR\ II_{p=0}$ is valid.
Using (\ref{iter}) for the definition of $U_4=1/{\cal Z}_4^{(1)}$
and $RR\ II_{p=0}$, one proves $RR\ I_{p=0}$. This
establishes the validity of both $RR\ I_{p=0}$ and $RR\ II_{p=0}$,
which is the starting point for the proof by induction.

Let us now assume that we have proved the validity of both
$RR\ I_p$ and $RR\ II_p$ for some given $p$. By using this induction
assumption and the definition (\ref{iter}), one then proves the
relation $RR\ II_{p+1}$ (\ref{recur2}). After that, by using the
definition (\ref{iter}) and $RR\ II_{p+1}$, one then proves $RR\
I_{p+1}$. This completes the induction step. We may thus conclude
that the relations $RR\ I_p$ and $RR\ II_p$ are valid for all $p\geq 0$.

One easily checks that the expressions
\be
U_j = \left( \frac{a(x)^{j+1} - a(x)^{-j-1}}{a(x)-a(x)^{-1}} \right) ^2
\label{recsol}
\ee
with
\be
a(x) = \half \left( \sqrt{\frac{1-3x}{x}} + \sqrt{\frac{1+x}{x}} \right)
\label{ax}
\ee
satisfy the recursion relations (\ref{recur1}) and the initial
conditions (\ref{init}). The function $F^2(x)$ is now expressed as a
convergent product
\be
F^2(x) = \prod_{m=2}^\infty \frac{ (U_{m}-1)^2}{U_m^2}
       = \prod_{m=2}^\infty \frac{ U_{m+1}U_{m-1} }{U_m^2}
       = \lim_{l \rightarrow \infty}
           \frac{U_1}{U_2} \frac{U_{l+1}}{U_l}
       = a(x)^2 x (x+1)  \,,
\label{f2x}
\ee
where we defined $U_1(x)=\frac{x+1}{x}$, in accord with
(\ref{recur1}), and where we used (\ref{recur1}) in the second equality. This
brings us to the following representation
of the number of regular Bethe Ansatz states with $M$ overturned spins
(using (\ref{xxx1}), (\ref{avonx0}), (\ref{avonx1}), (\ref{avonx2}),
(\ref{fx}), (\ref{ax}) and (\ref{f2x}) )
\bea
\lefteqn{
   \begin{array}{cc} \str & \\
   \displaystyle{ \sum_{M_m=0}^\infty } & n(\{ M_m \}) \\
   \str \scriptstyle{M= \sum_{m=1}^\infty m M_m} &  \end{array} =}
\nonu
&& = \frac{1}{2\pi i} \oint \frac{dx}{x^{M+1}} (1+x)^{N-M}
     \left( \frac{ (1+x) + \sqrt{(1+x)(1-3x)} }{2} \right)^{2M-N-1}
\nonu
&& = \frac{1}{2\pi i} \oint \frac{2\, dy}{y^{M+1}}
     \left( 2\,(1+y) \right)^{N-M}
     \left( 1+y + \sqrt{1-y^2} \right)^{2M-N-1} \,,
\eea
where the contour is a small circle around the origin and we used
the substitution $y=\frac{2x}{1-x}$.
Calling $y^{-1}= \cosh \phi$, the integral reduces to
$I_+-I_-$, where
\be
I_{\pm} = \frac{1}{2 \pi} \int_0^{2 \pi} d \varphi \,
   e^{- \phi (N-M \mp 1)} (1+e^{ \phi})^{N-1}
   = \left( \begin{array}{c} N-1 \\ N-M \mp 1 \end{array} \right)
\ee
where we wrote $\phi=\Lambda-i\varphi$, with $\Lambda \rightarrow \infty$.
This finally establishes the result (\ref{xxx1}) since
\bea
  \begin{array}{cc} \str & \\
  \displaystyle{ \sum_{M_m=0}^\infty } & n(\{ M_m \}) \\
  \str \scriptstyle{M= \sum_{m=1}^\infty m M_m} &  \end{array}
&=& \left( \begin{array}{c} N-1 \\ N-M - 1 \end{array} \right)
    - \left( \begin{array}{c} N-1 \\ N-M + 1 \end{array} \right)
\nonu
  &=& \left( \begin{array}{c} N \\ M \end{array} \right)
    - \left( \begin{array}{c} N \\ M-1 \end{array} \right) \,.
\eea

\vspace{5mm}
We still have to prove (\ref{xxx2}), which can be done as follows :
\bea
\sum_{M=0}^{[N/2]}&&\hskip -20pt
  \left( \left(\begin{array}{c} N \\ M \end{array} \right)
         - \left( \begin{array}{c} N \\ M-1 \end{array} \right)
  \right) (N - 2 \, M +1)\  =
\nonu
&=&  \sum_{M=0}^{[N/2]}\left( \begin{array}{c} N \\ M
\end{array} \right)(N - 2 \, M +1)-
\sum_{M=0}^{[N/2]-1}\left( \begin{array}{c} N \\ M
\end{array} \right)(N - 2 \, M -1)
\nonu
&=& 2 \sum_{M=0}^{[N/2]-1}\left( \begin{array}{c} N \\ M
\end{array} \right)+\left( \begin{array}{c} N \\ {\left\lbrack N\over
2\right\rbrack}\end{array} \right)\left(
1+N-2\left\lbrack\frac{N}{2}\right\rbrack\right)
\nonu
&=& \sum_{M=0}^{[N/2]-1}\left( \begin{array}{c} N \\ M
\end{array} \right)+\left( \begin{array}{c} N \\ {\left\lbrack N\over
2\right\rbrack}\end{array} \right)\left(
1+N-2\left\lbrack\frac{N}{2}\right\rbrack\right)
+\sum_{N-[N/2-1]}^{N}\left( \begin{array}{c} N \\ M
\end{array} \right)
\nonu
&=& \sum_{M=0}^{N}\left( \begin{array}{c} N \\ M
\end{array} \right)= 2^N\quad .
\label{last}
\eea
This completes the proof of equation (\ref{xxx2}).
\vspace{5mm}

\noindent {\sc STEP 2.}

\vspace{5mm}

The total number of states that are obtained from the SO(4) extended
Bethe Ansatz for the Hubbard model is given by (\ref{number}) and
(\ref{sum}).
The summations over the multiplicities $M_m$ and over the difference
$N-M$ in the summation (\ref{sum}) are precisely of
the type (\ref{xxx1}) and (\ref{xxx2}), respectively, if we substitute
$M \rightarrow \half(M_e-N+M)$ and $N \rightarrow M_e$. (Under these
summations the total number of electrons $N_e$ is kept fixed.)
The summation that remains after this `spin summation' is
\bea
\lefteqn{\# \; ({\rm eigenstates}) = \sum_{N_e=0}^{L} \; (L-N_e+1) }
\nonu
&& \hskip -35pt \begin{array}{cc}
     \times\left[ \STR \right. \;\;\;
     \displaystyle{\sum_{M_e=0}^{N_e} \sum_{M_m^\prime=0}^\infty} &
     \hskip -20pt2^{M_e}
     \left( \begin{array}{c} L \\ M_e \end{array} \right)
     \displaystyle{\prod_{n=1}^{\infty}} \left( \begin{array}{c}
     \str L-N_e + \sum_{m=1}^\infty (2m-t_{nm}) M^\prime_m \\
     \str M_n^\prime       \end{array} \right)
     \left. \STR \right]\,, \\
     \scriptstyle{\hspace{4mm} N_e = M_e + 2 \, \sum_{m=1}^\infty m
	          M^\prime_m} &
   \end{array}
\nonu &&
\label{sum2}
\eea
where as before $t_{nm} = 2 \,{\rm Min}(n,m) - \de_{nm}$.
In the next step we perform the summation with respect to all
$M_n^\prime$'s, using a similar kind of `summation device' as in STEP 1.
As a consequence of (\ref{binom}) we have
\be
(1-x^2)^{-1-B}(1+2x)^L
= \sum_{M_1^\prime=0}^\infty\sum_{p=0}^L \left( \begin{array}{c} B +M_1^\prime
   \\ M_1^\prime  \end{array} \right) \left( \begin{array}{c}
   L \\ p \end{array} \right)2^p\ x^{2 M_1^\prime +p}  \,,
\label{new1}
\ee
and therefore
\be
\frac{1}{2 \pi i} \oint \frac{dx}{x^{\gamma +1}} (1-x^2)^{-1-B}(1+2x)^L
= \sum_{M_1^\prime=0}^\infty\left( \begin{array}{c} B +M_1^\prime
   \\ M_1^\prime  \end{array} \right) \left( \begin{array}{c}
   L \\ \gamma -2 M_1^\prime \end{array} \right)2^{\gamma -2 M_1^\prime } \,.
\label{new2}
\ee
The integration is along a small contour around zero.
Defining $E=L-N_e$, $\gamma = N_e - 2\sum_{m=2}^\infty m M_m^\prime$
and $B =E+2\sum_{m=2}^\infty (m-1) M_m^\prime$, the r.h.s. of
(\ref{new2}) becomes the summation over $M_1^\prime$ in (\ref{sum2}),
if we solve the constraint in the sum in (\ref{sum2}) for $M_e=
N_e-2\sum_{m=1}^\infty m M_m^\prime$.
Using (\ref{new2}) in (\ref{sum2}) we then obtain the following
expression for the number of eigenstates :
\be
\# \; ({\rm eigenstates}) =
\frac{1}{2\pi i} \oint \frac{dx}{x^{L+1}(1-x^2)} (1+2x)^L
  \sum_{E=0}^L (E+1) \frac{x^E}{(1-x^2)^E} F(x)  \,,
\label{new3}
\ee
where

\be
F(x)=\sum_{{M_m^\prime}=0\atop m\ge 2}^\infty\prod_{n=2}^\infty \left(
\begin{array}{c} E+M_n^\prime +2 \sum_{m=n+1}^\infty (m-n) M^\prime_m
\\ \\
M_n^\prime\end{array} \right)\prod_{m=2}^\infty \left({\cal
Z}_m^{(0)}\right)^{M_m^\prime}\,,
\label{new4}
\ee
and
\be
{\cal Z}_m^{(0)}= \frac{x^{2m}}{(1-x^2)^{2(m-1)}} \,.
\ee
The summations over $M_2^\prime$, $M_3^\prime$,... have precisely the
form of the l.h.s. of (\ref{binom}) and can thus be performed easily.
The result is
\be
F(x) = \prod_{m=2}^\infty (1-{\cal Z}_m^{(m-2)})^{-1-E}\,,
\ee
where
\be
{\cal Z}_m^{(p)}=\frac{{\cal Z}_m^{(p-1)}}{(1-{\cal
Z}_{p+1}^{(p-1)})^{2(m-p-1)}}\,.
\ee
It can be shown along the lines given in STEP 1, that the quantities
$U_m={1\over{\cal Z}_m^{(m-2)}}$ obey the
recursion relation
\be
(U_{p+2}-1)^2 = U_{p+3} U_{p+1} \,, \quad p\geq 0
\label{recur1new}
\ee
with initial conditions
\be
U_1 = x^{-2},\qquad U_2 = \frac{(1-x^2)^2}{x^4}\,.
\label{initnew}
\ee
Equation (\ref{recur2}) is replaced by
\be
\frac{ {\cal Z}_{n+1}^{(p)}}{ {\cal Z}_{n}^{(p)}} =
\frac{ U_{p+1} }{ U_{p+2} } \,, \quad p\geq 0, \;\;  n\geq p+2 \,.
\label{recur2new}
\ee
Equation (\ref{new3}) now can be written as
\be
\# \; ({\rm eigenstates}) =
\frac{1}{2\pi i} \oint \frac{dx}{x^{L+1}} (1+2x)^L
  \sum_{E=0}^L (E+1) x^E [f(x)]^{-E-1}  \,,
\label{inth}
\ee
where
\be
f(x) = \prod_{l=1}^\infty (1-U_l^{-1})\,.
\label{eff}
\ee
The solution of the recursion relation (\ref{recur1new}) is again of the
form (\ref{recsol}), {\sl i.e.,}
$U_j = \left( \frac{a(x)^{j+1} - a(x)^{-j-1}}{a(x)-a(x)^{-1}} \right)
^2$, where now
\be
a(x) = \frac{1}{2x}+\sqrt{\frac{1}{4x^2}-1}
\ee
due to the new initial conditions (\ref{initnew}).
Insertion of the resulting expression for $U_l$ into (\ref{eff}) leads
to the following result for the function $f(x)$:
\be
2f(x) = 1 + \sqrt{1-4x^2}=2x\ a(x)\,.
\ee
Equation (\ref{inth}) can now be rewritten as
\be
\# \; ({\rm eigenstates}) = \sum_{E=0}^L (E+1)\ I(E)\,,
\label{states}
\ee
where
\be
I(E)=\frac{1}{2\pi i} \oint d\left(-\frac{1}{x}\right)\ (\frac{1}{x}+2)^L
\left\lbrack a(x)\right\rbrack^{-E-1}  \,.
\ee
The contour integration can be worked out as in section 2.
Defining $\alpha =\Lambda -i\varphi$ with $\Lambda \gg 1$ and
substituting $x=\frac{1}{e^{\alpha}+e^{-\alpha}}$ we obtain
\be
I(E)=I_+(E)-I_-(E)\,,
\label{eeee}
\ee
where
\be
I_\pm(E)=\frac{1}{2\pi}\int_0^{2\pi}d\varphi\
e^{\pm\alpha}\ e^{-(1+E)\alpha}\ \left(e^{\alpha\over
2}+e^{-{\alpha\over 2}}\right)^{2L}\,.
\ee
Expanding
\be
\left(e^{\alpha\over
2}+e^{-{\alpha\over 2}}\right)^{2L}=\sum_{p=0}^{2L}
\left(\begin{array}{c} 2L \\ p \end{array} \right)
e^{\alpha (L-p)}
\ee
and then using
\be
\frac{1}{2\pi}\int_{0}^{2\pi}d\varphi\ e^{\pm
i(n\varphi)}=\delta_{n,0}
\ee
in the resulting expression, we find that
\be
I_+(E)=\left(\begin{array}{c} 2L \\ L-E \end{array} \right),\qquad
I_-(E)=\left(\begin{array}{c} 2L \\ L-E-2 \end{array} \right)\,.
\ee
Plugging these results into (\ref{eeee}) and then (\ref{states}) we
are left with only a single summation
\be
\# \; ({\rm eigenstates}) = \sum_{E=0}^L (E+1)
\left\lbrace\left(\begin{array}{c} 2L \\ L-E \end{array} \right)
-\left(\begin{array}{c} 2L \\ L-E-2 \end{array} \right)\right\rbrace\,.
\ee
This summation can be performed the same way as (\ref{last}) and we
finally obtain the desired result
\be
\# \; ({\rm eigenstates}) = 4^L \,.
\ee
This concludes our two-step evaluation of the sum (\ref{sum}).

Using the above, we can obtain a closed expression for the number of
regular Bethe Ansatz states for given numbers $M$ and $N$ of spin-down
and spin-up electrons:
\bea
\lefteqn{ \!\!
   \begin{array}{cc}
   \displaystyle{ \sum_{M_e=0}^\infty \sum_{M_m=0}^\infty
        \sum_{M_m^\prime=0}^\infty } &
   n(M_e, \{ M_m \} , \{ M_m^\prime \}) \; = \\
   \strut  \scriptstyle{N_e = M_e + 2\, \sum m M^\prime_m} & \\
   \scriptstyle{M = \sum m (M_m + M_m^\prime)} &
   \end{array}}
\nonu
&& \!\! \left( \! \begin{array}{c} L \\ N \end{array} \! \right) \!
   \left( \left( \! \begin{array}{c} L \\ M \end{array} \! \right)
          + \left( \! \begin{array}{c} L \\ M-2 \end{array} \! \right) \right)
 - \left( \left( \! \begin{array}{c} L \\ N+1 \end{array} \! \right)
          + \left( \! \begin{array}{c} L \\ N-1 \end{array} \! \right) \right)
 \! \left( \! \begin{array}{c} L \\ M-1 \end{array} \! \right) .
\nonu &&
\eea
This formula is the close analogue of the result (\ref{xxx1})
for the XXX Heisenberg model.

We repeat once more our conclusion, which is that the combination
of the nested Bethe Ansatz with the $SO(4)$ symmetry of the
one-dimensional Hubbard model leads to a complete set of $4^L$
independent eigenstates.

It is a pleasure to thank C.N.~Yang for proposing the ideas
worked out in this paper. We thank S.~Dasmahapatra
for stimulating discussions. This work was supported by NSF Grant
PHY- 9107261.

\vspace{5mm}


\newpage\null
\appendix
\def\Lam#1{\Lambda_{#1}}

\section{Bound states in the one-dimensional Hubbard model}
\setcounter{equation}{0}

In this appendix we investigate the nature of the $\Lambda$ and $k-\Lambda$
strings in the one-dimensional Hubbard model. We show for explicit
examples, that both kinds of strings lead to certain kinds of bound
states ({\it i.e.}, the wave function decays exponentially with repect
to the differences of coordinates) .

\subsection{$\Lambda$ strings}
Let us consider the example of $N$ electrons with spin up and two electrons
with spin down forming a $\Lambda$ string, {\it i.e.},
\be
\Lambda_2={\Lambda_1}^* \quad , \qquad k_1,...,k_{N+2} {\ \rm real}\quad .
\ee
The set $\{\Lam{1},\Lam{2}|k_1,...,k_{N+2}\}$ must fulfill the periodic
boundary conditions (\ref{pbc}).

The wave function corresponding to this set of spectral parameters is
given by
\be
\psi_{_{-1,-1,1,...,1}}(x_{_1},x_{_2},...,x_{_{N+2}}) = \hskip
-10pt\sum_{P\in S_{2+N}}\hskip -8pt sgn(Q)  \ sgn(P) \
e^{i\sum_{j=1}^{2+N}k_{P_j}x_{Q_j}}\ \varphi(y_1,y_2|P) \quad ,
\label{B2}
\ee
with amplitudes
\bea
\varphi(y_1,y_2|P) &=& A_{id}\ F_P(\Lam{1},y_1)\ F_P(\Lam{2},y_2)\ +\
A_{21}\ F_P(\Lam{2},y_1)\ F_P(\Lam{1},y_2)
\nonu
&=&  A_{id}  \left(\ \prod_{i=1}^{y_1-1}\ {e^{(1)}_-(P_i)\over
     e^{(1)}_+(P_i)} \ \right) \ {1\over e^{(1)}_+(P_{y_1})}
     \left(\ \prod_{i=1}^{y_2-1}\ {e^{(2)}_-(P_i)\over
     e^{(2)}_+(P_i)} \ \right) \ {1\over e^{(2)}_+(P_{y_2})}
\nonu
&& + A_{21}  \left(\ \prod_{i=1}^{y_2-1}\ {e^{(1)}_-(P_i)\over
     e^{(1)}_+(P_i)} \ \right) \ {1\over e^{(1)}_+(P_{y_2})}
     \left(\ \prod_{i=1}^{y_1-1}\ {e^{(2)}_-(P_i)\over
     e^{(2)}_+(P_i)} \ \right) \ {1\over e^{(2)}_+(P_{y_1})}\quad .
\nonu &&
\label{B3}
\eea

We want to show that this wave function decays exponentially with respect
to the difference of the coordinates $y_1$ and $y_2$. The only nontrivial
$y$ dependent part of the wave function are the amplitudes $\varphi$.
Therefore it is sufficient to prove that they decay exponentially.

Taking exponentially small corrections $\Delta$ into account, the
$\Lam{}$ string is of the form
\bea
&& \Lam{1}= \Lam{} + {u+\Delta\over 2}
\nonu
&& \Lam{2}= \Lam{} - {u+\Delta\over 2}
   \qquad {\rm with\ } \Delta^*=-\Delta\qquad .
\eea
As all momenta $k_j$ are real this leads to the following inequalities
\be
\left| {e^{(1)}_-(i)\over e^{(1)}_+(i)} \right| >1\quad ,\qquad
\left| {e^{(2)}_-(i)\over e^{(2)}_+(i)} \right| <1\quad .
\label{inequ}
\ee
Using the periodic boundary conditions
\be
\prod_{i=1}^{N+2} {e^{(1)}_-(i)\over e^{(1)}_+(i)} =
{\Lam{2}-\Lam{1}-u\over\Lam{2}-\Lam{1} +u}= {A_{id}\over A_{21}}
\ee
we can express $A_{21}$ in terms of $A_{id}$.
Using the second set of periodic boundary conditions
\be
e^{i k_jL} = {e^{(1)}_-(j)\over e^{(1)}_+(j)}\
{e^{(2)}_-(j)\over e^{(2)}_+(j)}
\ee
we can express products over $e^{(\alpha)}_{\pm}$ in terms of
exponential factors of magnitude $1$.
Straightforward computations yield
\bea
\lefteqn{ \varphi(y_1,y_2|P) =
A_{id}\  e^{i\sum_{l=1}^{y_1-1}k_{P_l}L}
\left(\ \prod_{i={y_1}}^{y_2-1}\ {e^{(2)}_-(P_i)\over
e^{(2)}_+(P_i)} \ \right) \ {1\over e^{(1)}_+(P_{y_1})} \ {1\over
e^{(2)}_+(P_{y_2})} + }
\nonu
&&+\ A_{id}\  e^{i\sum_{l=1}^{y_2-1}k_{P_l}L}
\left(\ \prod_{i=1}^{y_1-1}\ {e^{(2)}_-(P_i)\over
e^{(2)}_+(P_i)} \ \right) \left(\ \prod_{i=y_2}^{N+2}\ {e^{(2)}_-(P_i)\over
e^{(2)}_+(P_i)} \ \right)
\ {1\over e^{(1)}_+(P_{y_2})} \ {1\over
e^{(2)}_+(P_{y_1})}\quad .
\nonu &&
\label{fi}
\eea

By definition $y_2 > y_1$ and the inequalities (\ref{inequ}) ensure that the
factor in brackets in the first term (and thus the whole term) in
(\ref{fi}) decays exponentially for $y_2 \gg y_1$.
The second term can be dropped, because for spin waves $N$ plays the
role of the lattice length, and in order to investigate asymptotic
properties of the wave function we should consider the infinite volume
limit, {\sl i.e.,} $N\longrightarrow\infty$. In this limit the second
term can be set to zero as $\left(\ \prod_{i=y_2}^{N+2}\
{e^{(2)}_-(P_i)\over e^{(2)}_+(P_i)} \ \right)$ vanishes.


\subsection{$k-\Lambda$ strings}

We consider the example $N=M=2$, {\it i.e.,} $2$ electrons with spin up
and $2$ with spin down. The periodic boundary conditions read
\bea
e^{i k_j L} &=& {e^{(1)}_-(j)\over e^{(1)}_+(j)} \
{e^{(2)}_-(j)\over e^{(2)}_+(j)}\qquad j=1,...,4
\nonu
\prod_{j=1}^4\ {e^{(\beta)}_-(j)\over e^{(\beta)}_+(j)} &=&
-\prod_{\alpha=1}^2{\Lam{\alpha}-\Lam{\beta}-u
\over\Lam{\alpha}-\Lam{\beta}+u}\quad\beta=1,2\qquad .
\label{per4}
\eea
A $k-\Lam{}$ string solution of these equations takes the following form
in the $L\longrightarrow\infty$ limit:
\bea
  \Lam{1} &=& \Lam{} -{u\over 2}=\Lam{2}^*
\nonu
  k_1 &=& \pi-\arcsin(\Lam{} -u)
\nonu
  k_2 &=& \arcsin(\Lam{})= \pi - k_3
\nonu
  k_4 &=& \pi-\arcsin(\Lam{} +u) \quad .
\eea
In the finite volume there exist two distinct configurations that both
lead to $k-\Lam{}$ strings in the limit $L\longrightarrow\infty$,
depending on whether $k_2$ and $k_3$ are real or complex for finite $L$.
\vskip .5cm
{\sl Case (i): $k_2$ and $k_3$ are complex}
\vskip .5cm
In this case the $k$'s can be rearranged such that
$Im(k_1)<Im(k_2)<0<Im(k_3)<Im(k_4)$ and the invariance of the periodic
boundary conditions under complex conjugation gives the additional
constraints
\be
k_3^*=k_2 \quad , \qquad k_4^*=k_1 \quad .
\ee
Taking this into account one obtains the following solution of
(\ref{per4}) for finite but large $L$:
\bea
&& \Lam{1} = \Lam{} -{u+\delta\over 2}=\Lam{2}^*
\nonu
&& sin(k_1) = \Lam{} -u-{\delta\over 2}+\epsilon_1=sin(k_4^*)
\nonu
&& sin(k_2) = \Lam{} +{\delta\over 2}+\epsilon_2=sin(k_3^*) \,,
\label{corra}
\eea
where $\delta$ is purely imaginary and $\epsilon_{1,2}$ are complex.
The exponentially small corrections are of the orders
\be
\delta={\cal O}\left(2u\ e^{-i(k_1+k_2^*)L}\right)\ , \quad
\epsilon_1={\cal O}\left(-2u\ e^{-ik_1L}\right)\ , \quad
\epsilon_2={\cal O}\left(-2u\ e^{-ik_1L}\right)\ .
\label{corr}
\ee
The wave function is given by (\ref{B2}) with $N=2$ and (\ref{B3}).
After inserting the values of the spectral parameters found in
(\ref{corra}) into (\ref{B2}) and (\ref{B3}) and re-normalising the
resulting expression one finds :
\bea
&& \psi_{_{-1,-1,1,1}}(x_1,x_2,x_3,x_4) =
\nonu
&& \!\!\!\! \left\{ \begin{array}{cc}
sgn(Q) (-1)^{y_1+y_2} \left( e^{i\sum_{j=1}^4 k_jx_{Q_j}}
-e^{i\sum_{j=1}^4 k_{R_j}x_{Q_j}} \right)\ & {\rm if}\
(y_1,y_2) \notin \left\{(1,2),(3,4)\right\} \\ \\
0 & {\rm else}\quad , \end{array} \right.
\nonu &&
\label{wavefn}
\eea
where $R$ is the permutation $(1,3,2,4)$.

Due to the ordering of the imaginary parts of the momenta $k_j$ both
terms on the l.h.s. of (\ref{wavefn}) decay exponentially with respect to
the magnitudes of differences of coordinates $|x_k-x_j|$ and thus the
wave function describes a bound state.

\vskip .5cm
{\sl Case (ii): $k_2$ and $k_3$ are real}
\vskip .5cm
In this case we must drop the constraint $k_2^*=k_3$. We then find the
following solution to (\ref{per4})
\bea
&& \Lam{1} = \Lam{} -{u+\delta\over 2}=\Lam{2}^*
\nonu
&& sin(k_1)= \Lam{} -u-{\delta\over 2}+\epsilon_1=sin(k_4^*)
\nonu
&& sin(k_2)= \Lam{} +\epsilon_2
\nonu
&& sin(k_3)= \Lam{} +\epsilon_3 \quad ,
\eea
where $\epsilon_1$ is complex and $\epsilon_{2,3}$ are real,
while $\delta$ is again purely imaginary.
The corrections are of the orders
\bea
\epsilon_1 &=& {\cal O} \left(-2u\ e^{-ik_1L}\right)
\nonu
 \delta &=& {\cal O} \left({4i\ Re(\epsilon_1)\over cot({k_2L\over 2})+
                   cot({k_3L\over 2})}\right)
\nonu
 \epsilon_2 &=& {\cal O} \left({-i\over 2}cot({k_2L\over 2})\delta\right)
\nonu
 \epsilon_3 &=& {\cal O} \left({-i\over 2}cot({k_3L\over 2})\delta\right)
                    \quad .
\label{corr2}
\eea

The computation of the wave function is analogous to case (i), the
only difference being a new re-normalisation constant.
The wave function is given by the same expression as in case (i).
Again it describes a bound state although $k_2$ and $k_3$ are now
real.
\setcounter{equation}{0}
\section{The $M=N=1$ sector in the Hubbard model}

In this appendix we further work out the stucture
of the Bethe Ansatz wavefunctions in the sector
$M=N=1$. In that sector the
wavefunctions depend on parameters
$k_1$, $k_2$ and $\La$. Our general analysis in sections 3
and 4 gives the following possibilities:
(i) we can have $M_e=2$, $M_1=1$, which gives real values
    for $k_1$, $k_2$ and $\La$, or
(ii) we can have $M_1^\prime=1$, which gives a $m=1$ $k$-$\La$-string
    with complex conjugate $k_1$, $k_2$ and real $\La$.
According to the counting of section 3, we expect to have $\half L(L-1)$
real solutions (i) and $L-1$ string solutions (ii), giving
a total number of $\half (L-1)(L+2)$ Bethe Ansatz  states in this
sector. However, we already mentioned that, in the
context of the XXX Heisenberg model, there is a `redistribution
phenomenon' between different types of solutions, which does not
affect the total number of states within a sector, but which does
affect the distribution of those states over various types of
string-solutions and real solutions
$^{\hbox{\scriptsize\cite{eks3}}}$. In this appendix we will
establish a very similar result for the Hubbard model: we will
find that the numbers of the real solutions (i) and string solutions
(ii) are not always given by the values quoted above,
but that the expected total number $\half (L-1)(L+2)$ of solutions
in this sector can be rigorously established.

Let us consider the Bethe equations for periodic boundary conditions
in the sector $M=N=1$
\bea &&
e^{i k_1 L} =
  \frac{\sin k_1 - \La - \frac{u}{2}}{\sin k_1 - \La + \frac{u}{2}}\ , \quad
e^{i k_2 L} =
  \frac{\sin k_2 - \La - \frac{u}{2}}{\sin k_2 - \La + \frac{u}{2}}
\nonu && \qquad
\prod_{i=1}^2
  \frac{\sin k_i - \La - \frac{u}{2}}{\sin k_i - \La + \frac{u}{2}} =1 \ .
\eea
Since we assume $\La$ to be finite, we can solve for it and find
$\La = \half(\sin k_1 + \sin k_2)$. The equations then reduce to
\be
e^{ik_1L} = \frac{ \sin k_1 - \sin k_2 -u}{ \sin k_1 - \sin k_2 +u}\ ,
\quad e^{i (k_1+k_2)L}=1 \ .
\label{eqx-1}
\ee
We can solve the second equation by putting $k_1+k_2=\frac{2 \pi}{L}m$,
with $m=0,1, \ldots,2L-1$ and write $k_1=\frac{\pi}{L}m +x$ and
$k_2=\frac{\pi}{L}m - x$. The remaining equation reads
\be
e^{i(\pi m + L x)} = \frac{-\frac{4}{U}\cos(\frac{\pi m}{L})
 \sin x - i}{-\frac{4}{U}\cos(\frac{\pi m}{L}) \sin x + i} \,.
\label{eqx}
\ee

One easily checks that, if $x_0$ solves this equation for $m=m_0$,
then $x=x_0+\pi$ solves the equation for $m=m_0+L$, and that the
resulting wavefunctions are the same. We can thus restrict our attention
to $m=0,1, \ldots L-1$ and $-\pi\leq x< \pi$.

Let us now try to find real solutions $x$ for the equation (\ref{eqx})
for given $m$. Taking a logarithm we have
\be
\arctan \left( -\frac{U}{4} \cos(\frac{\pi m}{L}) \sin x \right)
= \half Lx -  \pi n \,,
\label{eqx-2}
\ee
where $n$ is an arbitrary integer for $m$ odd and half an odd
integer for $m$ even. It is rather straightforward to
solve this equations by a graphical method: one plots both
the l.h.s\ and the r.h.s.\ (for various $n$) of these equations on
the interval $-\pi \leq x <\pi$ and reads of the intersection points,
which are then solutions of the equation. For large enough $|U|$
this procedure is easily carried out and one finds the following.

For $m$ even there are solutions for
$n=-\frac{L+1}{2}, \ldots, \frac{L-1}{2}$, which are $L$
solutions in total. Since solutions $x$ and $-x$ are equivalent
(and $x=0$ is not among the solutions) we should divide this
number by 2. Using that there are $L/2$ possible even values for $m$
we thus find $L^2/4$ solutions.
For $m$ odd one finds non-equivalent solutions for
$n=1,2, \ldots, \frac{L-1}{2}$ for each $m$, which gives
a total number of $(L^2-2L)/4$. (The solutions $x=0$, $x=\pm\pi$,
which exist for generic $U$, give vanishing wavefunctions
in general.) Adding up the contributions from odd and even $m$,
we find $\half L (L-1)$, which is indeed the number predicted
by the counting in section 3.

However, let us now assume that $m$ is odd and that $U$ is close
to a critical value $U_m$, which we define by
\be
\frac{U_m}{4} \cos(\frac{\pi m}{L}) = \frac{L}{2} \,.
\ee
At the value $U=U_m$ the curve for the l.h.s.\ of (\ref{eqx-2})
has slope $L/2$ at $x=\pm \pi$, and at the value $U=-U_m$ the curve
for the l.h.s.\ of (\ref{eqx-2}) has slope $L/2$ at $x=0$.
Since the r.h.s.\ is given by straight lines of slope $L/2$,
and since both curves already had intersections at
$x=0$ (for $n=0$) and $x=\pm\pi$ (for $n=\pm L/2$)
(which did however not give rise to non-trivial wavefunctions),
it will be clear that the number of intersections changes when $U$
reaches the critical values $\pm U_m$. In fact, one finds one extra real
solution $x$ (together with the equivalent solution $-x$)
for a given odd $m$ as soon as $U < |U_m|$. For example, if $U$ is
such that $U_1>U>U_3>....>0$ there will be one extra real solution
to the equations (\ref{eqx-1}).

\vspace{6mm}

The complex values for $x$ which solve the equation (\ref{eqx})
are of the form $x=iy$ or $x=\pi+iy$ with $y$ real. In a way
similar to what we showed above, one can analyse the equations
for the real quantity $y$ by a graphical method.
If $|U|$ is sufficiently large, one finds precisely one complex
solution for $m=1,2, \ldots, L$ with the exception of $m=L/2$.
In that case, there are thus $L-1$ complex solutions, which is
in agreement with the counting of section 3.

However, from the graphical analysis one finds that the complex
solution for a given odd $m$ disappears as soon as $|U|$ is chosen
to be smaller than $U_m$. Note that this happens precisely in the
regime where we have found one extra real solution!

We thus find a redistribution phenomenon, where solutions change
their nature as a function of $U$, in analogy to what we found
for the XXX Heisenberg model in \cite{eks3}. In the Hubbard model the
phenomenon is easily understood: if $|U|$ is made small enough,
the interactions become so weak that some of the bound states
(with complex $x$) decay into real solutions (with real $x$).

When $|U|$ is chosen to be equal to one of the critical values
$U_m$, there do exist nontrivial wavefunctions with
$x=0$ or $x=\pm \pi$, {\it i.e.}, with coinciding $k_1$ and $k_2$.
These wavefunctions can be seen to be nonvanishing by a
renormalisation \`a la l'H\^opital.

In all cases, the total number of eigenfunctions of the
hamiltonian in the sector $M=N=1$ is found to be
$\half (L-1)(L+2)$, which is the value predicted by the
counting in section 2, and used for the proof of completeness
in section 3.

\end{document}